\documentclass[preprint, aip, pof,showpacs]{revtex4-1} % fleqn
\usepackage{graphicx, bm, bbm,amsmath, amssymb, epsfig,multirow,tabularx}

\begin{document}

% \DeclareFixedFont{\BBb}{U}{msb}{m}{n}{1\@ptsize}

\newcommand{\bq}{\begin{equation}}
\newcommand{\eq}{\end{equation}}
\newcommand{\bqs}{\begin{equation*}}
\newcommand{\eqs}{\end{equation*}}
\newcommand{\bqa}{\begin{eqnarray}}
\newcommand{\eqa}{\end{eqnarray}}
\newcommand{\bqas}{\begin{eqnarray*}}
\newcommand{\eqas}{\end{eqnarray*}}
\def\etal{{\em et al.\ }}
\newcommand{\dtt}{\partial_{tt}}
\newcommand{\ds}{\partial_{s}}
\newcommand{\dt}{\partial_t}
\newcommand{\z}{\zeta}
\newcommand{\Dfig}[2]{\includegraphics*[width=#2in]{#1.eps}}
\newcommand{\hs}{\hat{s}}
\newcommand{\hn}{\hat{n}}
\newcommand{\hy}{\hat{y}}
\newcommand{\hp}{\hat{[p]}}
\newcommand{\hv}{\hat{V}}
\newcommand{\hg}{\hat{\gamma}}
\newcommand{\bM}{\bar{M}}
\newcommand{\bU}{\bar{U}}
\newcommand{\bV}{\bar{V}}
\newcommand{\bfr}{\bar{f}}
\newcommand{\bbeta}{\bar{\beta}_M}
\newcommand{\br}{\bar{r}}
\newcommand{\bl}{\bar{l}}
\def\Xint#1{\mathchoice
   {\XXint\displaystyle\textstyle{#1}}%
   {\XXint\textstyle\scriptstyle{#1}}%
   {\XXint\scriptstyle\scriptscriptstyle{#1}}%
   {\XXint\scriptscriptstyle\scriptscriptstyle{#1}}%
   \!\int}
\def\XXint#1#2#3{{\setbox0=\hbox{$#1{#2#3}{\int}$}
     \vcenter{\hbox{$#2#3$}}\kern-.5\wd0}}
\def\ddashint{\Xint=}
\def\dashint{\Xint-}
\newenvironment{Eonefigs}[3]
{
        \begin{figure} [ht]
          \begin{center}
          \begin{tabular}{c}
              \Dfig{#1}{#3} \\
           \vspace{-.25in}
          \end{tabular}
         \caption{{\footnotesize #2}}
          \label{#1}
          \end{center}
        \vspace{-0in}
       \end{figure}
}{}
\newenvironment{Etwofigs}[5]
{
        \begin{figure} [htpb]
          \begin{center}
          \setlength{\tabcolsep}{.1in}
          \begin{tabular}{cc}
          \multicolumn{1}{l}{$\bf{(a)}$}&\multicolumn{1}{l}{$\bf{(b)}$}\\
              \Dfig{#1}{#4} & \Dfig{#2}{#5} \\
                   \vspace{-.25in}
          \end{tabular}
          \caption{\footnotesize #3}
          \label{#1}\label{#2}
          \end{center}
        \vspace{-.25in}
        \end{figure}
}{}

\newenvironment{Etwofigss}[5]
{
\begin{figure} [htpb]
          \begin{center}
          \setlength{\tabcolsep}{.1in}
          \begin{tabular}{cc}
          \multicolumn{1}{l}{$\bf{(a)}$}&\multicolumn{1}{l}{$\bf{(b)}$}\\
              \Dfig{#1}{#4} &\vspace{0.3in} \Dfig{#2}{#5} \\
                   \vspace{-.25in}
          \end{tabular}
          \caption{\footnotesize #3}
          \label{#1}\label{#2}
          \end{center}
        \vspace{-.25in}
        \end{figure}
}{}

\newenvironment{Ethreefigs}[7]
{
        \begin{figure} [htpb]
          \begin{center}
          \setlength{\tabcolsep}{.1in}
          \begin{tabular}{ccc}
          \multicolumn{1}{l}{$\bf{(a)}$}&\multicolumn{1}{l}{$\bf{(b)}$}&\multicolumn{1}{l}{$\bf{(c)}$}\\
              \Dfig{#1}{#5} & \Dfig{#2}{#6}&\Dfig{#3}{#7} \\
                    \vspace{-.25in}
          \end{tabular}
          \caption{\footnotesize #4}
          \label{#1}\label{#2}\label{#3}
          \end{center}
        \vspace{-.25in}
        \end{figure}
}{}
\newenvironment{Ethreefigsv}[7]
{
        \begin{figure} [htpb]
          \begin{center}
          \setlength{\tabcolsep}{.1in}
          \begin{tabular}{cc}
          \multicolumn{1}{l}{$\bf{(a)}$}&\multicolumn{1}{l}{$\bf{(b)}$}\\
              \Dfig{#1}{#5} & \Dfig{#2}{#6} \\
          \multicolumn{2}{l}{$\bf{(c)}$}\\
          \multicolumn{2}{c}{\Dfig{#3}{#7}}
                    \vspace{-.25in}
          \end{tabular}
          \caption{\footnotesize #4}
          \label{#1}\label{#2}\label{#3}
          \end{center}
        \vspace{-.25in}
        \end{figure}
}{}
\newenvironment{Ethreefigsvp}[7]
{
        \begin{figure} [htpb]
          \begin{center}
          \setlength{\tabcolsep}{.1in}
          \begin{tabular}{c}
          \multicolumn{1}{l}{$\bf{(a)}$}\\
          \Dfig{#1}{#5}\\
          \multicolumn{1}{l}{$\bf{(b)}$}\\
           \Dfig{#2}{#6}\\
          \multicolumn{1}{l}{$\bf{(c)}$}\\
          \Dfig{#3}{#7}
                    \vspace{-.25in}
          \end{tabular}
          \caption{\footnotesize #4}
          \label{#1}\label{#2}\label{#3}
          \end{center}
        \vspace{-.25in}
        \end{figure}
}{}
\newenvironment{Efourfigsv}[9]
{
        \begin{figure} [htpb]
          \begin{center}
           \setlength{\tabcolsep}{.1in}
          \begin{tabular}{cc}
          \multicolumn{1}{l}{$\bf{(a)}$}&\multicolumn{1}{l}{$\bf{(b)}$}\\
              \Dfig{#1}{#6} & \Dfig{#2}{#7} \\
              \multicolumn{1}{l}{$\bf{(c)}$}&\multicolumn{1}{l}{$\bf{(d)}$}\\
                     \Dfig{#3}{#8} & \Dfig{#4}{#9} \\
                   \vspace{-.25in}
          \end{tabular}
          \caption{{\footnotesize #5}}
          \label{#1}\label{#2}\label{#3}\label{#4}
          \end{center}
        \vspace{-.25in}
        \end{figure}
}{}
\newenvironment{Efourfigs}[7]
{
        \begin{figure} [htpb]
          \begin{center}
           \setlength{\tabcolsep}{.1in}
          \begin{tabular}{cc}
          \multicolumn{1}{l}{$\bf{(a)}$}&\multicolumn{1}{l}{$\bf{(b)}$}\\
              \Dfig{#1}{#6} & \Dfig{#2}{#6} \\
              \multicolumn{1}{l}{$\bf{(c)}$}&\multicolumn{1}{l}{$\bf{(d)}$}\\
                     \Dfig{#3}{#7} & \Dfig{#4}{#7} \\
                   \vspace{-.25in}
          \end{tabular}
          \caption{{\footnotesize #5}}
          \label{#1}\label{#2}\label{#3}\label{#4}
          \end{center}
        \vspace{-.25in}
        \end{figure}
}{}
\newenvironment{Efivefigs}[8]
{
        \begin{figure} [htpb]
          \begin{center}
          \setlength{\tabcolsep}{.1in}
          \begin{tabular}{cc}
          \multicolumn{1}{l}{$\bf{(a)}$}&\multicolumn{1}{l}{$\bf{(b)}$}\\
              \Dfig{#1}{#7} & \Dfig{#2}{#7} \\
            \multicolumn{1}{l}{$\bf{(c)}$}&\multicolumn{1}{l}{$\bf{(d)}$}\\
              \Dfig{#3}{#7} & \Dfig{#4}{#7} \\
              \multicolumn{2}{l}{$\bf{(e)}$}\\
            \multicolumn{2}{c}{\Dfig{#5}{#8}}\\
                    \vspace{-.25in}
          \end{tabular}
          \caption{{\footnotesize #6}}
          \label{#1}\label{#2}\label{#3}\label{#4}\label{#5}
          \end{center}
        \vspace{-.25in}
        \end{figure}
}{}

\newenvironment{Efivefigsv}[8]
{
        \begin{figure} [htpb]
          \begin{center}
          \setlength{\tabcolsep}{.1in}
          \begin{tabular}{cccc}
          \multicolumn{4}{l}{$\bf{(a)}$}\\
  \multicolumn{4}{c}{\Dfig{#1}{#7}}\\
\multicolumn{1}{l}{$\bf{(b)}$}& \multicolumn{1}{l}{$\bf{(c)}$}&\multicolumn{1}{l}{$\bf{(d)}$}& \multicolumn{1}{l}{$\bf{(e)}$}\\
             \Dfig{#2}{#8} & \Dfig{#3}{#8}& \Dfig{#4}{#8}&\Dfig{#5}{#8}\\
                    \vspace{-.25in}
          \end{tabular}
          \caption{{\footnotesize #6}}
          \label{#1}\label{#2}\label{#3}\label{#4}\label{#5}
          \end{center}
        \vspace{-.25in}
        \end{figure}
}{}
\newenvironment{Esixfigs}[8]
{
        \begin{figure} [htpb]
          \begin{center}
          \setlength{\tabcolsep}{.1in}
          \begin{tabular}{ccc}
          \multicolumn{1}{l}{$\bf{(a)}$}&\multicolumn{1}{l}{$\bf{(b)}$}&\multicolumn{1}{l}{$\bf{(c)}$}\\
              \Dfig{#1}{#8} & \Dfig{#2}{#8}&\Dfig{#3}{#8} \\
            \multicolumn{1}{l}{$\bf{(d)}$}&\multicolumn{1}{l}{$\bf{(e)}$}&\multicolumn{1}{l}{$\bf{(f)}$}\\
              \Dfig{#4}{#8} & \Dfig{#5}{#8}&\Dfig{#6}{#8} \\
                    \vspace{-.25in}
          \end{tabular}
          \caption{{\footnotesize #7}}
          \label{#1}\label{#2}\label{#3}\label{#4}\label{#5}\label{#6}
          \end{center}
        \vspace{-.25in}
        \end{figure}
}{}
\newenvironment{Esixfigsv}[8]
{
        \begin{figure} [htpb]
          \begin{center}
          \setlength{\tabcolsep}{.1in}
          \begin{tabular}{cc}
          \multicolumn{1}{l}{$\bf{(a)}$}&\multicolumn{1}{l}{$\bf{(b)}$}\\
              \Dfig{#1}{#8} & \Dfig{#2}{#8}\\
              \multicolumn{1}{l}{$\bf{(c)}$}&\multicolumn{1}{l}{$\bf{(d)}$}\\
                \Dfig{#3}{#8}& \Dfig{#4}{#8} \\ \multicolumn{1}{l}{$\bf{(e)}$}&\multicolumn{1}{l}{$\bf{(f)}$}\\
              \Dfig{#5}{#8}&\Dfig{#6}{#8}
                    \vspace{-.25in}
          \end{tabular}
          \caption{{\footnotesize #7}}
          \label{#1}\label{#2}\label{#3}\label{#4}\label{#5}\label{#6}
          \end{center}
        \vspace{-.25in}
        \end{figure}
}{}
\def\Xint#1{\mathchoice
{\XXint\displaystyle\textstyle{#1}}%
{\XXint\textstyle\scriptstyle{#1}}%
{\XXint\scriptstyle\scriptscriptstyle{#1}}%
{\XXint\scriptscriptstyle\scriptscriptstyle{#1}}%
\!\int}
\def\XXint#1#2#3{{\setbox0=\hbox{$#1{#2#3}{\int}$}
\vcenter{\hbox{$#2#3$}}\kern-.5\wd0}}
\def\ddashint{\Xint=}
\def\dashint{\Xint-}

\title{Stability and scalability of piezoelectric flags}
\author{Xiaolin Wang, Silas Alben}
\email{wxiaolin@umich.edu, alben@umich.edu}
\affiliation{Department of Mathematics, University of Michigan}
\author{Chenyang Li, Yin Lu Young}
\email{licheny@umich.edu, ylyoung@umich.edu}
\affiliation{Department of Naval Architecture and Marine Engineering, University of Michigan}

\date{\today}

\begin{abstract}
We investigate 
the effect of piezoelectric (PZT) material on the flutter speed, vibration mode and frequency, and energy harvesting power and efficiency of a flexible flag in various fluids. 
We develop a fully coupled fluid-solid-electric model by combining the inviscid vortex sheet model with a
linear electro-mechanical coupling model. A resistance only circuit and a resonant resistance-inductance (RL) circuit are considered. 
 For a purely resistive circuit, an increased electro-mechanical coupling factor results in an increased flutter speed, vibration frequency, 
averaged electric power and efficiency. 
A consistent optimal resistance is found that maximizes the flutter speed and the energy harvesting power. 
For a resonant RL circuit, by tuning the inductance to match
 the circuit frequency to the flag's vibration frequency, the flutter speed can be greatly decreased, and a larger averaged power and efficiency are obtained.
 We also consider a model scale set-up with several commonly used commercial materials for operating in air and water. 
Typical ranges of dimensionless parameters are obtained for four types of material that span a wide range of solid density and rigidity values. 
We find that the resistance only circuit is more effective when the flag is placed in a lighter fluid (e.g. air), while the RL circuit is able to 
reduce the flutter speed when the flag is placed in a heavier fluid (e.g. water). 
\end{abstract}
\maketitle
\section{Introduction}
Piezoelectric (PZT) material has drawn enormous attention in recent decades due to its ability
 to convert mechanical kinetic energy into electrical potential energy, and vice versa.
It has been applied to energy harvesting devices to power micro sensors/actuators \cite{anton2007review,cook2008powering}, and to passively control structural vibrations
\cite{hagood1991damping,behrens2003broadband,corr2002comparison}. The fundamental
 mechanism in the electro-mechanical process requires vibrations of the solid, which can be realized through natural flow-induced vibrations.

A flexible elastic flag undergoes flow-induced vibration through the competition between the destabilizing effect of the fluid pressure and the
 stabilizing effect of the flag's internal and bending rigidity. In particular,
 the flag will develop large-amplitude deformations if the flow speed is above a certain critical value, which we define as flutter.
Much previous work has been conducted on this phenomenon
through experiments \cite{taneda1968waving,huang1995flutter, zhang2000flexible,watanabe2002experimental, tang2003flutter,shelley2005heavy}, 
inviscid and viscous numerical simulations
\cite{guo2000stability,zhu2002simulation,argentina2005fluid,connell2007flapping,alben2008flappingstate,tang2007instability,michelin2008vortex,akcabay2012hydroelastic}, 
and linear stability analysis \cite{shelley2005heavy,watanabe2002theoretical,eloy2007flutter}. The critical flutter speed was found over a large range of mass ratio (ratio
of solid to fluid inertial force), and increases as the mass ratio decreases. The effect of mass ratio on the flutter mode was observed in experiments
 and numerical simulations, and both found
that the flag flutters in higher modes when the mass ratio is small and the first mode shape is never unstable for a cantilever plate \cite{guo2000stability, eloy2007flutter}. Post-flutter behavior was also discussed in previous work, and
flags were found to transition from limit-cycle oscillations
to more chaotic states as the fluid velocity increases beyong flutter \cite{patil2001limit,alben2008flappingstate, virot2013fluttering}.

The objective of this work is to consider the effect of the piezoelectric material on the flutter speed, vibration mode and frequency, and energy harvesting power and efficiency of a flexible flag.
Previous studies found that the electro-mechanical coupling effect of the piezoelectric material acts to increase the bending stiffness of the system,
 and therefore will tend to stabilize the flag by increasing the critical flutter speed \cite{akcabay2012hydroelastic,doare2011piezoelectric}. 
The output circuit also has an important impact on the flutter boundary. It is often described as a shunting circuit as the piezoelectric material is
modeled as a current source in parallel with its internal capacitance \cite{ottman2002adaptive,guan2007efficiencies}, and the output circuit provides extra paths for the current.  
A resistive shunting circuit with a single resistor and a resonance shunting circuit with resistor-inductor (RL) are two commonly used output circuits for energy harvesting and
vibration control \cite{hagood1991damping}. With the resistive shunting, Akcabay and Young \cite{akcabay2012hydroelastic} studied the fluid-solid interaction
of a bimorph piezoelectric cantilevered beam in a viscous flow using the immersed boundary method, and they found an optimal resistance value for energy harvesting
 efficiency. Michelin and Doar\'{e} \cite{doare2011piezoelectric} studied the piezoelectric coupling effect on the local and global instabilities, as well as on energy conversion
efficiency for a flexible plate using the double wake inviscid model. They found that the critical flutter speed increases as the piezoelectric coupling increases,
and the energy conversion efficiency depends on piezoelectric coupling, loaded resistance, and the thickness ratio of the PZT patch to the substrate material.
In another work, they studied bodies of small span using slender body theory\cite{michelin2013energy}.
The increase of the flutter speed by the PZT material has an important impact on vibration control.
On the other hand, higher effective system stiffness also indicates the flutter initiates at a larger flow speed, which will impact the energy harvesting potential and efficiency. In one aspect, this limits the effectiveness of the piezoelectric material in energy harvesting,
especially in water and other heavier fluids because the required flutter speed is higher when the solid-to-fluid mass ratio is smaller.
Some recent work suggests that this difficulty may be overcome by connecting the PZT flag to a well-tuned resonant shunting circuit.
Xia \etal \cite{xia2015fluid} found that the flag's vibrating frequency can lock-in to the circuit resonant frequency for a parallel connected RL circuit with a flapping PZT flag using the slender body
theory. They showed a significantly enhanced energy efficiency and a large decrease of the flutter speed in the lock-in regime. Similar results were obtained
for the passive control of vibrating plates. Li \etal \cite{li2014Passive} investigated the effect of the RL circuit on the passive control of a hydrofoil. They also
observed a maximum damping enhancement for an open-loop vibration control application via PZT with an RL circuit when the inductor is tuned to match the circuit resonant frequency with the foil resonant frequency.

In this work, we extend the vortex-sheet model by Alben \cite{alben2009simulating}
 and include the piezoelectric effect by addition of a linear electro-mechanical coupling model to develop a fully-coupled model to
predict the electro-fluid-solid response of a piezoelectric flag connected with an output electric circuit in incompressible and inviscid flow.
In this model,
the vortex layers induced by the body tend to sheets of infinitesimal thickness. A vortex-sheet model is efficient for slightly viscous flow because it can present the flow using
vortex sheet, which is only one-dimensional in a 2D flow. It allows flow separations at prescribed locations (usually sharp edges), but it agrees quantitatively with Navier-Stokes models in predicting the strength
of shed circulation, and the location of the primary vortices in the wake of an oscillated plate \cite{sheng2012simulating}.
 Previous work also shows a good agreement of the vortex-sheet model in predicting the stability boundary of a flapping flag \cite{alben2008flappingstate}.
 Linear electro-mechanical coupling models were applied to cantilever flags for a single-patch PZT layer model\cite{erturk2009experimentally,dutoit2007experimental}
 and a multiple-patch layer \cite{bisegna2006optimized,doare2011piezoelectric} model, and were found to compare well with experimental results at low to
moderate strains. We then derive the important dimensionless parameters that govern the model, and investigate their effects on the critical flutter speed, vibration
frequency, mode shapes and energy harvesting power and efficiency with the numerical simulation results. Next, we
discuss the effectiveness and scalability of bimorph PZT flags for energy harvesting
and vibration control applications in air versus in water and for different substrate materials.

The paper is organized as follows: Sec. II describes the vortex-sheet-plate model for the piezoelectric flag;
 Sec. III studies the effects of piezoelectric parameters on the stability boundary; Sec. IV presents the scalability results for actual devices with different materials
 and fluids. The conclusions are presented in Sec. V.

\section{Modelling}
\subsection{Plate-Vortex-Sheet Model}
We consider here the motion of a cantilevered thin plate or beam in an inviscid flow. The beam is of chord length $C$, span length $S$,
mass per unit chordwise length $m_s$, and bending rigidity $B$. The flow and the motion of the beam is assumed to be 2D, so all the properties are
 uniform in the spanwise direction. The thickness of the beam is assumed to be much smaller than its
chord and span, so we model it as a 1D inextensible elastic sheet. Piezoelectric layers are combined in parallel
on both sides of the beam by poling the layers in the same direction \cite{erturk2009experimentally,wang1999constitutive}.
The beam moves under the pressure forces of a surrounding inviscid and incompressible flow of density (mass per unit volume) $\rho_f$ with
 an uniform upstream horizontal velocity $U$. A schematic figure of the piezoelectric beam is shown in figure \ref{fig:schematic}(a).
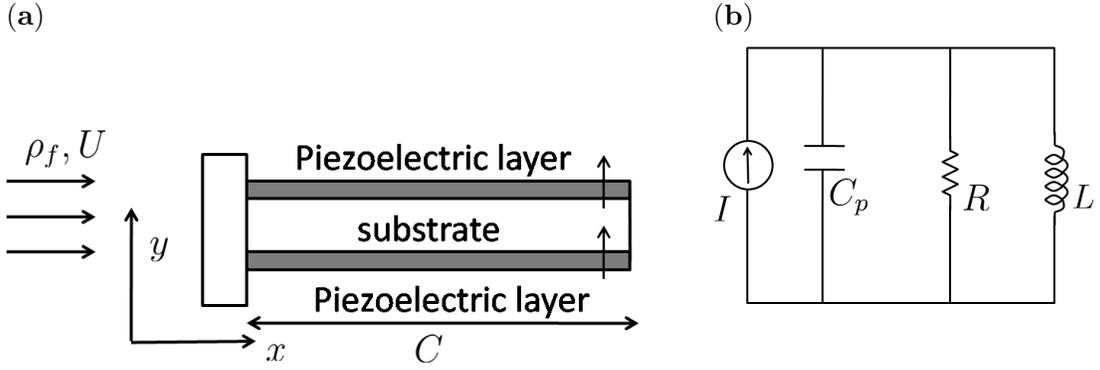
\Etwofigs{schematic}{circuit}{\label{fig:schematic}(a). A cantilever beam of chord length $C$ subject to a uniform flow of speed $U$ and density $\rho_f$. The arrows on the piezoelectric layer
indicate the two layers are poled in the same direction.
 (b). Equivalent circuit parallel connected with a resistor. $R$ is the resistance, $L$ is the inductance, $C_p$ is the equivalent capacitance of the PZT material,
 $I$ is the current generated by electro-mechanical conversion effect. }{3.5}{2}

The instantaneous position of the beam is described as $\z(s,t)=x(s,t)+iy(s,t)$, where $s$ is the arclength. Assuming an Euler-Bernoulli model for the plate, the
bending moment at a given position consists of an internal rigidity of the material and the piezoelectric coupling:
\bq
M(s,t)=B\kappa(s,t)-\beta_MV(s,t)\label{eq:bendmoment}
\eq
where $B=EI$ is the bending rigidity with $E$ the material's Young's modulus and $I$ its second moment of area, $\kappa(s,t)$ is the curvature of the beam,
 $\beta_M$ is a electro-mechanical coupling factor indicating the bending moment generated per unit voltage,
 and $V(s,t)$ is the electric potential generated
between the two piezoelectric layers. Therefore, the governing equation for $\z$ is:
\bq
m_s\partial_{tt}\z(s,t)=\partial_s(T(s,t)\hat{s})-B\partial_s(\partial_s\kappa(s,t)\hat{n})+\beta_M\partial_s(\partial_s V(s,t)\hat{n})-S[p](s,t)\hat{n}.
\label{eq:beameq}
\eq
Here, $T(s,t)$ is the beam's tension force accounting for its inextensibility, and $[p](s,t)$ is the pressure jump across the beam.
 $\hs=\partial_s\zeta$ and $\hn=i\hs$ represent the unit tangent
and normal vectors to the beam, respectively.
 The piezoelectric composite beam is parallel connected with a resistance-inductance (RL) output circuit. The voltage $V(s,t)$ is governed by the 
linear dynamic equation for the circuit:
\bq
c_p\displaystyle\partial_{tt}V+\frac{\partial_t V}{r}+\frac{V}{l}+\beta_M\partial_{tt}\kappa=0\label{eq:constitutive}
\eq
where $c_p=\displaystyle\frac{C_p}{C}$, $r=RC$ and $l=LC$ are the surface density of the internal capacitance, resistance, and inductance, 
respectively with $C$ as the
chord length of the beam.
 The equation is derived by representing the piezoelectric element
as a current source in parallel with its internal capacitance \cite{ottman2002adaptive,guan2007efficiencies} and has been used before in previous work by Xia \etal\cite{xia2015fluid}. The equivalent circuit is shown in figure \ref{fig:schematic}(b). Since the piezoelectric layer is parallel connected, the value of capacitance is twice that on each layer.
The piezoelectric effect also plays a role in the trailing edge condition, where the pressure induced moment and forces are zero:
\bq
T=\partial_s\kappa=0,\quad \kappa=\displaystyle\frac{\beta_M}{B}V
\eq

 If we assume the voltage $V=V(t)$ is constant
along the piezoelectric layer, then the above equation can be rewritten by integrating along the arclength:
\bq
C_p\displaystyle\partial_{tt}V+\frac{\partial_t V}{R}+\frac{V}{L}+\beta_M\int_0^C\partial_{tt}\kappa(s',t)ds'=0
\eq
In this formulation,
$\beta_M\partial_s(\partial_s V(s,t)\hat{n})=0$, equation (\ref{eq:beameq}) then reduces to the 
Bernoulli beam equation\cite{alben2009simulating}, i.e., without piezoelectric effect. In this case, the piezoelectric material affects the flag only through the boundary condition,
and we can view the piezoelectric material as adding an extra concentrated moment at the free tip of the beam.

The pressure forces are computed by the inviscid vortex-sheet model described by Alben \cite{alben2009simulating}.
This model allows flow separation at sharp edges only, but it agrees quantitatively with Navier-Stokes models in predicting the strength of shed circulation, and the location of the primary vortices in the wake of an oscillating plate \cite{sheng2012simulating}. In the vortex-sheet model, the flow is computed in terms of the
position and strength of a single vortex sheet. The vortex sheet consists of a ``bound'' part ($C_b$) on the beam, and it separates from the beam's trailing edge into
a ``free'' part ($C_f$). The position of the vortex sheet is denoted $\z(s,t)$, which is the same as the beam's position when $0\leq s\leq C$. The strength
of the vortex sheet is denoted $\gamma(s,t)$, which is also the jump in tangential velocity of the beam. On the bound vortex sheet,
the flow satisfies the no-penetration condition, which requires the body's normal velocity to equal the flow's normal velocity:
\bq
Re(\hn\partial_t\z^*)=Re\left(\hn\left(U+\frac{1}{2\pi i}\dashint_{C_b}\frac{\gamma(s',t)}{\z(s,t)-\z(s',t)}ds'+b_\delta(s,t)\right)\right),\,\z(s,t)\in C_b\label{eq:nopenetration}
\eq
Here, the symbol ``$*$'' denotes complex conjugate and $U$ is the speed of the background flow. The special integral symbol in the equation denotes a principal-value integral, and the general solution $\gamma(s,t)$ has inverse-square-root singularities at the beam's leading
and trailing edge \cite{muskhelishvili2008singular}. The last term is a regularized Biot-Savart integral using the smoothing parameter $\delta$ of Krasny's method
\cite{krasny1986desingularization}:
\bq
b_\delta(s,t)=\displaystyle\frac{1}{2\pi i}\int_{C_f} \frac{{\z^*(s,t)-\z^*(s',t)}}{|\z(s,t)-\z(s',t)|^2+\delta(s',t)^2}ds',\,\z(s,t)\in C_f
\eq
where
\bq
\delta(s,t)=\delta_0\left(1-\exp\left(-|\z(L,t)-\z(s,t)|^2/4\delta_0^2\right)\right)
\eq
with $\delta_0=0.2$. This tapered regularization decreases the effect of the regularization term at the trailing edge as $\delta(s,t)\rightarrow 0$ with $s\rightarrow C$, and it allows smooth vorticity and velocity dynamics for the free vortex sheet \cite{alben2010regularizing}.

On the free sheet $C_f$ in the wake, the vortex sheet evolves according to the Birkhoff-Rott equation \cite{birkhoff1962helmholtz,rott1956diffraction}:
\bq
\frac{\partial\z^*}{\partial t}(s,t)=U+\frac{1}{2\pi i}\dashint_{C_b}\frac{\gamma(s',t)}{\z(s,t)-\z(s',t)}ds'+b_\delta(s,t),\quad \z(s,t)\in C_f\label{eq:b-r}
\eq
The pressure jump $[p](s,t)$ across the beam is related to the vortex sheet strength $\gamma(s,t)$ using a version of the unsteady Bernoulli equation \cite{hou2001boundary,alben2012attraction,jones2003separated}:
\bq
\gamma_t(s,t)+\partial_s((\tau_f(s,t)-\tau_s(s,t))\gamma(s,t))=\frac{1}{\rho_f}\partial_s[p](s,t)\label{eq:pressure}
\eq
Here, $\tau_f(s,t)$ and $\tau_s(s,t)$ are the tangential velocities of the fluid and the solid respectively. We couple equations (\ref{eq:beameq}), (\ref{eq:constitutive}), (\ref{eq:nopenetration}), (\ref{eq:b-r}) and (\ref{eq:pressure}) together and solve the nonlinear system with a numerical method similar to Alben's work \cite{alben2009simulating}.
We compute the free vortex sheet with an explicit method and solve for the unknown variables including beam position, 
vortex sheet strength, pressure and voltage using Broyden's method \cite{ralston2012first}. 
More detailed discussion can be found in previous work\cite{alben2009simulating}, including how the vorticity in the free vortex sheet is generated at the trailing edge using the Kutta condition, and how the Cheybyshev-Lobatto nodes are implemented to improve the convergence of integral-type equation.

\subsection{Nondimensionalization}
We nondimensionalize the governing equations by: the semi-chord $b=\displaystyle\frac{1}{2}C$,
 the fluid density $\rho_f$, and the internal capacitance density $c_p$ as the characteristic length, density and capacitance density, respectively. We choose the velocity scale as $u=\sqrt{\displaystyle\frac{B}{\rho_fb^3S}}$ so that
the dimensionless inflow velocity is $\bar{U}=U\sqrt{\displaystyle\frac{\rho_fb^3S}{B}}$ and
$\bU^2$ represents the ratio of the fluid kinetic energy to the solid elastic energy. Therefore, we have the characteristic time as
$\tau=\displaystyle\frac{b}{u}=\displaystyle\sqrt{\frac{\rho_fb^5S}{B}}$, and the following dimensionless variables:
\bq
\bar{\z}=\frac{\z}{b}, \quad\bar{t}=\frac{t}{\tau},\quad\bar{\kappa}=\kappa b, \quad\bar{T}=T\frac{b^2}{B},\quad\bar{p}=p\frac{b^3S}{B},
\quad \bar{V}=V\sqrt{\frac{c_pb^2}{B}},\quad \bar{f}=f\sqrt{\frac{\rho_fb^5S}{B}}\nonumber
\eq
Here, $f$ is the beam vibration frequency.
We list the dimensionless parameters and the corresponding physical meanings in
 Table \ref{tab:nondimen}.
\begin{table}[ht]
\caption{\label{tab:nondimen}Dimensionless parameters governing the piezoelectric flexible beam}
\begin{tabular}{p{4.8cm}lcp{6.7cm}}
\hline\hline
Dimensionless group& & &Physical meaning\\
\hline
Mass ratio&$\quad \bar{M}$&$\quad\displaystyle\frac{m_s}{\rho_fbS}$&Ratio of the solid inertia to fluid inertia\\ \hline
Reduced velocity&$\quad \bar{U}$&$\quad U\sqrt{\displaystyle\frac{\rho_fb^3S}{B}}$&
Ratio of the fluid kinetic energy to the solid elastic energy\\ \hline
Piezoelectric/Mechanical coupling factor&$\quad\bar{\beta}_M$&$\quad \beta_M\sqrt{\displaystyle\frac{1}{Bc_p}}$
&Ratio of the stored electrical energy in PZT to the solid elastic energy\\ \hline
Reduced resistance density&$\quad \bar{r}$&$\quad rc_p\sqrt{\displaystyle\frac{B}{\rho_fb^5S}}$&Ratio of the produced energy dissipated in resistance and stored as elastic energy to fluid kinetic energy\\ \hline
Reduced inductance density&$\quad \bar{l}$&$\quad \displaystyle\frac{lc_pB}{\rho_fb^5S}$&Ratio of the produced energy dissipated in inductance and stored as elastic energy to fluid kinetic energy\\ \hline\hline
\end{tabular}
\end{table}

We then obtain the following dimensionless equations:
\bq
\bar{M}\partial_{\bar{t}\bar{t}}\bar{\z}(\bar{s},\bar{t})=\partial_{\bar{s}}(\bar{T}\hs)-\partial_{\bar{s}}(\partial_{\bar{s}}\bar{\kappa}\hn)+\bbeta\partial_{\bar{s}}
(\partial_{\bar{s}} \bar{V}\hn)-[\bar{p}]\hn
\eq
\bq
\partial_{\bar{t}\bar{t}}\bar{V}+\frac{\partial_{\bar{t}} \bar{V}}{\bar{r}}+\frac{\bar{V}}{\bar{l}}+\bar{\beta}_M\partial_{\bar{t}\bar{t}}\bar{\kappa}=0
\eq

\section{Results}
\subsection{Validation of flutter velocity and frequency without piezoelectric effect}
We first briefly review the flutter boundary for flags in an unbounded flow without the piezoelectric effect.
The flag deformation grows exponentially when the speed of the background flow exceeds the flutter speed \cite{shelley2005heavy}.
The flutter boundary curve gives the parameter values at which the net damping of the beam becomes zero. In the simulation, it is obtained by imposing small initial
deflection and tracking the deflection's growth or decay in time.
Below the curve that represents the flutter boundary (in $\bU$), the initial perturbation on the beam will decay in time to zero due to the net positive damping; above the curve,
 small perturbations will grow exponentially due to net negative damping.

In figure \ref{fig:FlutterBdNoPiezo} (a), we show the reduced flutter speed $\bU$ for beams for a large range of mass ratios $\bM$
together with the previous experimental,
 numerical and analytical results, where all are obtained without the piezoelectric effect.
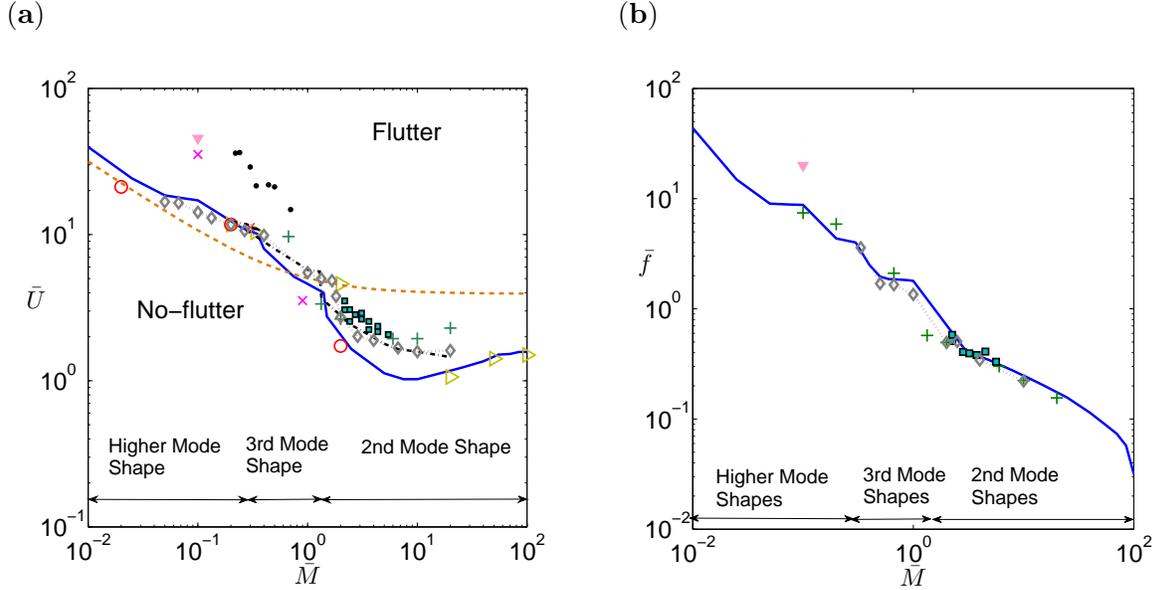
 \Etwofigs{FlutterBdNoPiezoComp}{FlutterFreqNoPiezoComp}{\label{fig:FlutterBdNoPiezo} (a) Flutter velocity $\bar{U}$
 at different mass ratio $\bar{M}$. (b) Response frequency $\bfr$ at the flutter boundary with different mass ratio $\bM$.
Current inviscid plate-vortex sheet method (solid line).
Experimental results by: Watanabe \etal \cite{watanabe2002experimental} conducted in air at $Re\approx100000$ ($\bullet$); Huang \cite{huang1995flutter} conducted in air at $Re\approx 40000$ ($\blacksquare$);
Shelley \etal \cite{shelley2005heavy} conducted in water at $Re\approx60000$ ($\blacktriangledown$). Numerical simulation
results of viscous flow model by: Akcabay and Young \cite{akcabay2012hydroelastic} at $Re=840$ ($+$); Connell and Yue \cite{connell2007flapping} at
$Re=1000$ ($\times$). Numerical simulation results of inviscid flow model by: Tang and Po\"\i doussis \cite{tang2007instability}
with potential flow theory (dotted-diamond line); Michelin \etal \cite{michelin2008vortex} with point vortex model ($\rhd$);
 Do\'{a}re and Michelin \cite{doare2011piezoelectric} with double wake model ($\bigcirc$). Inviscid analytical results by: Shelley \etal \cite{shelley2005heavy} with
infinite flag model (dashed line); Eloy \etal \cite{eloy2007flutter} with double wake model and linear stability analysis (dash-dot line).
}
{3}{3}

In general, our inviscid vortex sheet model consistently underestimates the critical flutter speed compared to the viscous and the experimental results.
 This is not only due to the viscous effect, but due to the fact that we neglect the material damping in the model.
Typical beams have a material damping coefficient of $1\%-5\%$ \cite{adams1999building}, and previous work shows that increases in material damping will lead to a delay in flutter, particularly for large $\bM$ \cite{akcabay2012hydroelastic}. The vortex sheet model agrees well with other inviscid model results, especially in the low mass ratio regime ($\bM<1$).
 When $\bM$ is large, the beam flutters at lower mode in the flapping state, and the assumption that its shape is close to a travelling wave in Shelley's infinite flag
 model \cite{shelley2005heavy} is violated. Therefore, we observe a large discrepancy between the solid line (current model) and the dashed line (Shelley's model) 
for $\bM>1$ in
 panel (a) if figure \ref{fig:FlutterBdNoPiezo}. We also notice differences in the flutter boundary between our model and the inviscid results from Eloy \etal \cite{eloy2007flutter}
and Tang and Po\"\i doussis \cite{tang2007instability}. Eloy \etal used a ``double wake'' model, which has an artificial upstream wake, while our model
only includes the wake shed from the trailing edge. This leads to different flow descriptions, especially for heavier beam $\bM>1$. Tang and Pa\"\i doussis used a vortex panel
method and modeled the vortex wake as a series of vortices whose strength depends on the Kutta-condition. They also included $0.4\%$ material damping in the model,
 which could be the reason that their flutter boundary curve is closer to the experimental results by Huang \cite{huang1995flutter}.

In panel (b) of figure \ref{fig:FlutterBdBeta}, we show the flapping frequency $\bfr$ of the beam at the flutter boundary. The velocity is chosen to be right above
 the flutter boundary according to panel (a). Our results agree well with previous numerical and experimental results.

In the flapping state,
the flag shows different mode structures which are determined by the most unstable mode. This phenomenon is observed in 
experiments \cite{eloy2007flutter,watanabe2002experimental}, simulations \cite{tang2007instability, michelin2008vortex, akcabay2012hydroelastic}
 and linear stability analysis \cite {watanabe2002theoretical,eloy2007flutter}. 
In figure \ref{fig:FlutterMode}, we show the typical mode shapes of the beam oscillations just above the flutter boundary. 
If we define the mode shape by the number of necks in the flags' envelops similarly to the work of Michelin \etal \cite{michelin2008vortex}, we obtain flapping
modes ranging from 2 (one neck) and 3 (two necks) to higher values depending on the mass ratio $\bM$. When $\bM$ is small, the relative dominance of fluid inertia and high beam flexibility
at the flutter boundary allow higher wavenumber bending. This observation is in agreement with previous work \cite{akcabay2012hydroelastic,watanabe2002experimental,michelin2008vortex}.
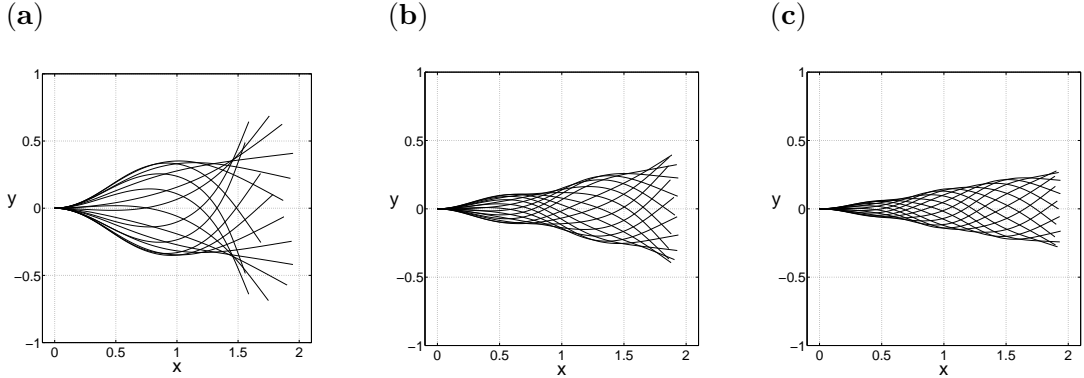
\Ethreefigs{ModeR1.4_U.1.4_R3.0_R4.0}{ModeR1.0.6_U.6_R3.0_R4.0}{ModeR1.0.2_U.12_R3.0_R4.0}{\label{fig:FlutterMode}Flag snapshots corresponding to different modes of beam oscillations just above the flutter boundary. (a).
Second mode shape, $\bM=4$ and $\bU=1.4$. (b). Third mode shape, $\bM=0.6$ and $\bU=6$. (c). Fourth mode shape, $\bM=0.2$ and $\bU=12$.}{1.8}{1.8}{1.8}
\subsection{Influence of piezoelectric material on flutter}
Now we consider the effect of the piezoelectric material on flutter. With the piezoelectric material, the effective system stiffness is increased by decreasing the 
fluid disturbing moment via the electro-mechanical coupling term with $\bbeta$ in equation
 (\ref{eq:bendmoment}). This term dissipates fluid kinetic energy through the output circuit. Overall, the piezoelectric material functions as a stabilizer
 to the system by increasing the effective stiffness and damping.
\subsubsection{Piezoelectric effects without inductance}
We first consider a simpler circuit without the inductor. In this case, the circuit equation becomes:
\bq
\partial_{\bar{t}} \bar{V}+\frac{\bar{V}}{\br}+\bbeta\partial_{\bar{t}}\bar{\kappa}=0\label{eq:constitutive2}
\eq
In figure \ref{fig:FlutterBdBeta}, we plot the critical flutter speed
and the corresponding response frequency right on the flutter boundary with a fixed value of $\br$
 and various values of the electric-mechanical coupling factor $\bbeta$ over a wide range of mass ratio $\bM$.
We define the critical flutter speed to be $U_0$ and the flutter frequency to be $f_0$
 when the piezoelectric material is not included, and plot them with solid lines in figure \ref{fig:FlutterBdBeta}.
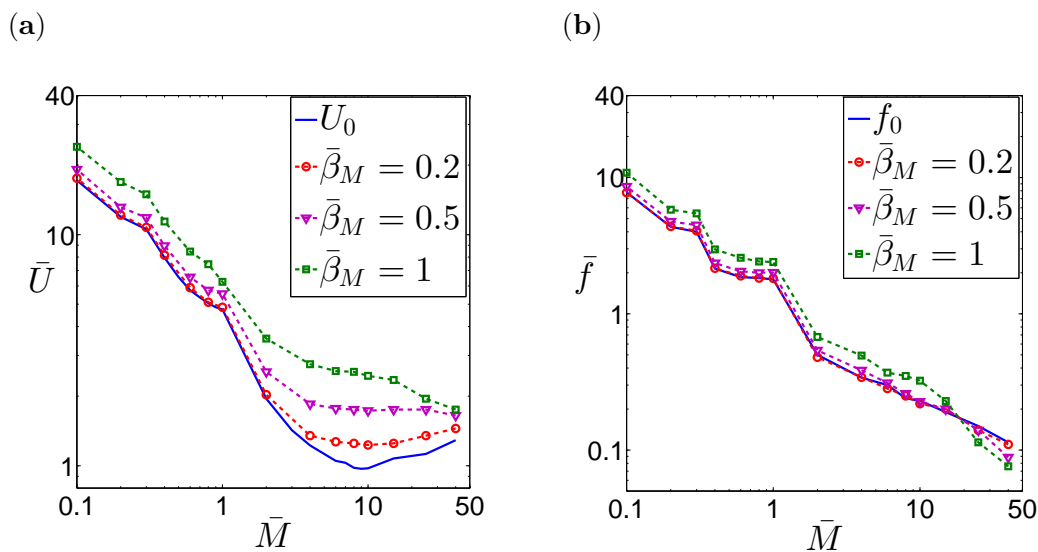
\Etwofigs{FlutterBoundPiezoBeta}{FlutterFreqBeta}{\label{fig:FlutterBdBeta} Reduced flutter speed (a) and frequency (b) results for fixed $\br=1$ and $\bbeta=0.2,0.5,1$.
The flutter boundary results without piezoelectric effect are plotted with solid line.}
{2.7}{2.7}
The critical flutter speed increases when $\bbeta$ increases for a given $\br$. The piezoelectric material
not only affects the flutter boundary, but also changes the vibration frequency at the flutter boundary.
 As the coupling factor $\bbeta$ increases, the vibration frequency increases as shown in figure \ref{fig:FlutterBdBeta}(b),
due to the increase in effective stiffness by the piezoelectric material, which will be shown later when discussing the effect of the resistance.

A flapping flag without the piezoelectric effect changes from a pre-flutter to a periodic flutter state to a chaotic state as the incoming flow
velocity increases \cite{alben2008flappingstate}. Similarly, with the piezoelectric effect, the flag undergoes periodic vibration within certain ranges of
flow velocity. In the periodic state, we define the amount of the electric power generated in the circuit by a time and space averaged power over one period as
$P_R=\left\langle\displaystyle\frac{\bar{V}^2}{\bar{R}}\right\rangle$. In figure \ref{fig:FlutterEnergyBeta}(a), we show the power
obtained right above the critical flutter speed $\bar{U}$ with a fixed value of $\br=1$ for varying $\bM$ and $\bbeta$.
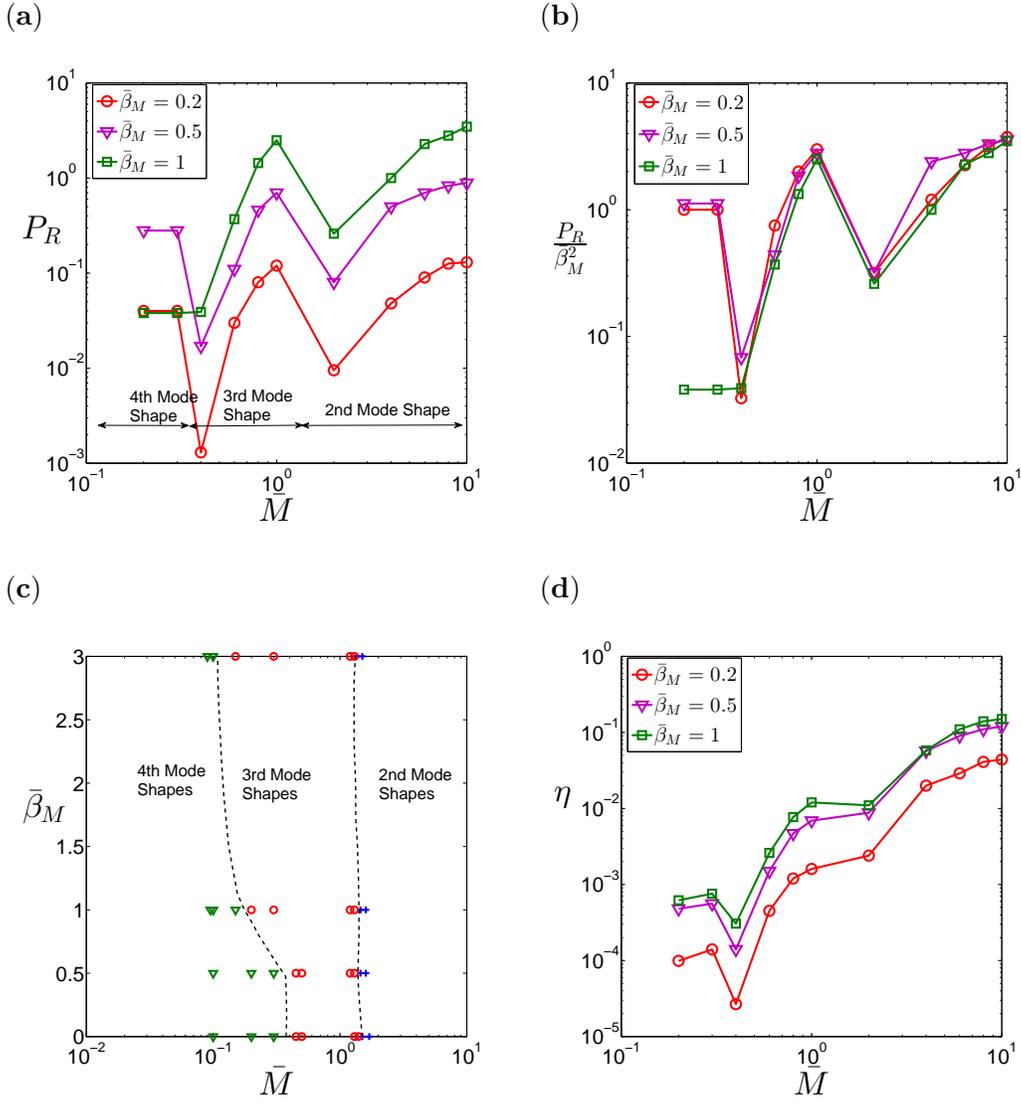
\Efourfigs{FlutterEnergyBeta}{FlutterEnergyBetaCollapse}{Modeshape_beta}{FlutterEfficiencyBeta}{\label{fig:FlutterEnergyBeta}(a). Averaged power $P_R$ vs. mass ratio
$\bM$ for fixed $\br=1$ and various $\bbeta=0.2,0.5,1$.
(b). The averaged power obtained with different $\bbeta$ are rescaled by $\bbeta^2$ and a good collapse is obtained.
(c). Diagram of mode shapes for different mass ratio $\bM$ and $\bbeta=0,0.5,1,3$. $\bbeta=0$ indicates the case without piezoelectric effect. (d).
The averaged efficiency $\eta$ vs. $\bM$ for for fixed $\br=1$ and various $\bbeta=0.2,0.5,1$. }{2.6}{2.6}
In general, for the same mass ratio, the power increases with $\bbeta$ due to higher electro-mechanical coupling, i.e., more fluid kinetic energy is converted 
to electric energy.
The power $P_R$ scales as $\bbeta^2$ as shown in figure
\ref{fig:FlutterEnergyBeta}(b), where we divide the power by $\bbeta^2$ and find a collapse of all the power curves. 
Equation ($\ref{eq:constitutive2}$) can be rewritten as $\bar{V}=\displaystyle\bar{\beta}_Me^{-\frac{\bar{t}}{\bar{r}}}\int e^{\frac{\bar{t}}{\bar{r}}}\partial_{\bar{t}}\bar{\kappa}d\bar{t}$. If we assume
the curvature only weakly depends on the piezoelectric coupling factor, then $\bar{V}$ scales with $\bbeta$ and the power scales with $\bbeta^2$.
As the mass ratio decreases, we obtain multiple peaks in $P_R$, and each of the peaks corresponds to a transition to a higher mode shape. A larger critical flutter
speed is required when the flag changes to a higher mode shape, and the higher speed leads to more fluid kinetic energy converted to electrical energy, and hence 
larger power.
The deviation in trend observed in figure \ref{fig:FlutterEnergyBeta}(a) for $\bbeta=1$ at $\bM=0.2$ and $0.3$ is because
the flag is still in the third mode shape for $\bbeta=1$, but changes to a fourth mode shape for $\bbeta=0.2$ and $0.5$. Therefore a higher energy is
 obtained at low $\bbeta$ values. In figure \ref{fig:FlutterEnergyBeta}(c), we show a diagram of different mode shapes obtained with various $\bM$ and $\bbeta$. 
For the same mass ratio, the increasing
$\bbeta$ increases the effective stiffness of the flag. Therefore, the flag oscillates in a lower mode shape, and the curve which differentiates different modes tends to lower
mass ratios as $\bbeta$ increases.

Similar to some previous work \cite{akcabay2012hydroelastic,xia2015fluid,akaydin2010energy}, we consider
the energy harvesting efficiency as the ratio of the output electric power to available fluid power:
\bq
\eta=\frac{\left\langle V^2/R \right\rangle}{\rho_fU^3AS}=\frac{P_R}{\bar{U}^3\bar{A}}
\eq
where $A$ and $\bar{A}$ are the dimensional and dimensionless maximum vibration amplitude of the flag during one period. 
We show the corresponding efficiency obtained with the same parameters in figure
\ref{fig:FlutterEnergyBeta}(d). In general, the efficiency decreases as $\bM$ decreases. Although the power increases when the mode shape changes,
the flutter speed $\bU$ also increases. Therefore, the value of efficiency, which scales as $\bU^{-3}$, depends on the competition of the power and the flutter speed, 
and can either becomes larger ($\bM=0.4$) or smaller ($\bM=1$) when the mode shape changes. 

 Next, we consider the effect of the resistance on the flutter boundary. In figure \ref{fig:FlutterBdR}, we show the flutter boundary results for a fixed value of $\bbeta=0.5$ and various $\br$.
\Efourfigs{FlutterBoundPiezoR1}{FlutterBoundPiezoR2}{FlutterFreqR1}{FlutterFreqR2}{\label{fig:FlutterBdR}
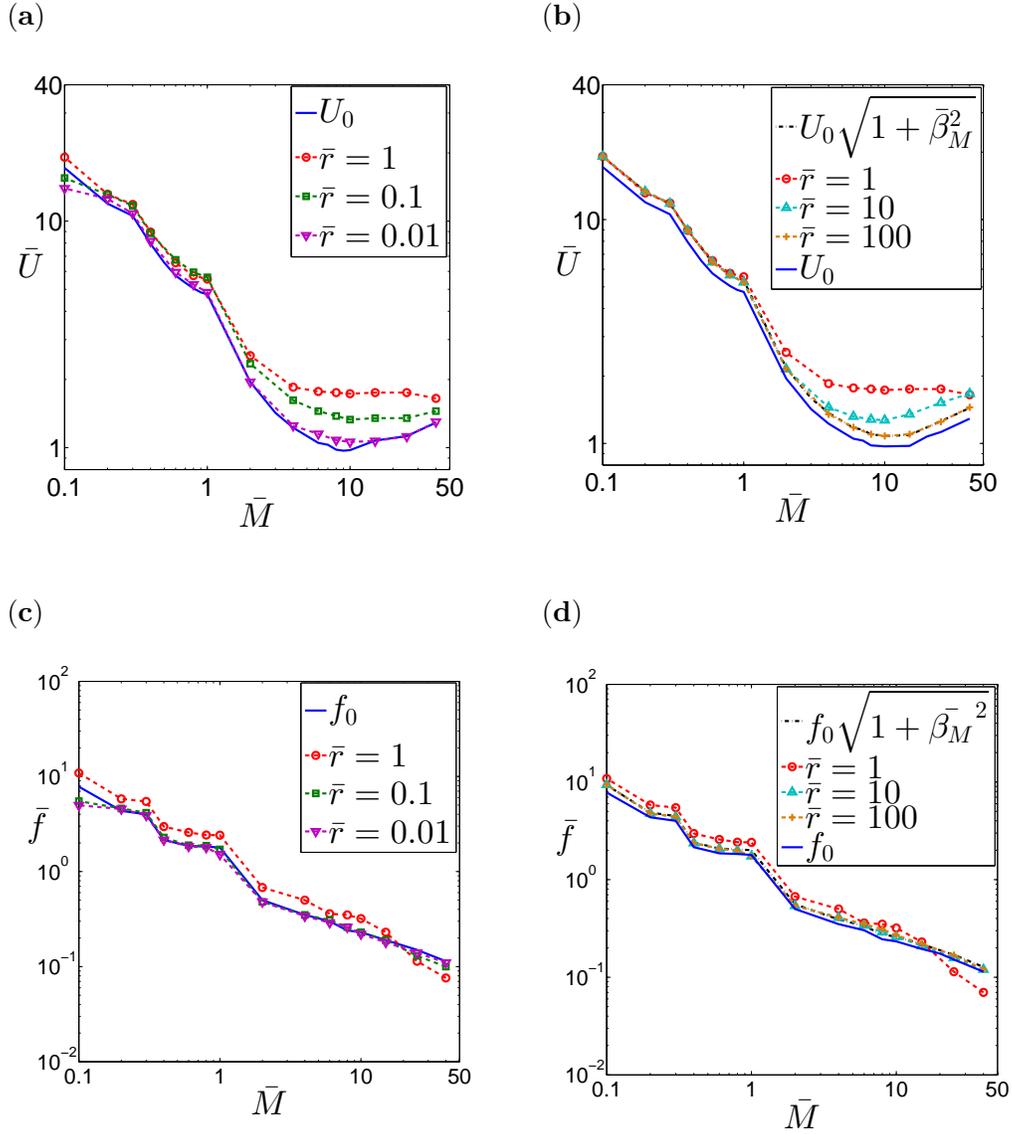
(a). Critical flutter speed for fixed $\bbeta=0.5$ and $\br=0.01,0.1,1$. The flutter speed without piezoelectric effect
is $U_0$ plotted with solid line. (b). Critical flutter speed for fixed $\bbeta=0.5$ and $\br=1,10,100$. Solid line shows $U_0$ and dash-dot line shows the curve
 $U_0\left(1+\bbeta^2\right)^{1/2}$. (c). Frequency values at the flutter boundary for fixed $\bbeta=0.5$ and $\br=0.01,0.1,1$. The frequency without piezoelectric effect $f_0$
is plotted with solid line. (d). Frequency values at the flutter boundary for fixed $\bbeta=0.5$ and $\br=1,10,100$.
Solid line shows $f_0$ and dashed-dot line shows
the curve $f_0\left(1+\bbeta^2\right)^{1/2}$.}{2.6}{2.6}
The dimensionless parameter $\br$ is a property of the electric circuit. When $\br$ is small,
the circuit is close to a short circuit and the effect of piezoelectric material decreases.
 As shown in panel (a), the critical flutter velocities are closer to $U_0$ as $\br\rightarrow 0$. When $\br$ is large, the circuit is close to an open circuit.
In the limiting case when $\br\rightarrow\infty$, the circuit equation can be rewritten as
\bq
\bV=-\bbeta\bar{\kappa}
\eq
therefore, the flag behaves equivalently to a flexible flag with a higher bending rigidity of $B(1+\bbeta^2)$ and 
 the critical velocity should be $U_0\left(1+\bbeta^2\right)^{1/2}$. The piezoelectric material therefore acts to increase the effective bending stiffness of
 the system. 
When the resistance has a moderate value, the voltage does not depend linearly on the curvature. However, it can still be represented by the product
of $\bbeta$ and a function of curvature, which suggests an increase in the system's bending stiffness with an increasing $\bbeta$.  
We notice that this observation is consistent with the results obtained by Doar\'{e} and Michelin
 \cite{doare2011piezoelectric}.
 In figure \ref{fig:FlutterBdR}(b), we plot the stability boundary results for fixed $\bbeta=0.5$ and $\br=1,10,100$ together with the curve $U_0\left(1+\bbeta^2\right)^{1/2}$ and $U_0$,
 and we can see that when $\br=100$, the curve approaches the curve $U_0\left(1+\bbeta^2\right)^{1/2}$.
 The corresponding vibration frequencies for varying $\br$ are shown in panels (c) and (d) of figure \ref{fig:FlutterBdBeta}. 
Similarly to the critical speed, when $\br\rightarrow 0$, the frequency tends to $f_0$; when $\br\rightarrow\infty$,
the frequency tends to $f_0\left(1+\bbeta^2\right)^{1/2}$.

Unlike the trend with varying $\bbeta$, the flutter velocity as well as the flutter frequency change non-monotonically with $\br$, as shown in figure \ref{fig:FlutterBdR}.
In figure \ref{fig:optR}, we plot the flutter velocity, the corresponding vibration frequency, the averaged electric power and efficiency right above the flutter boundary
varying over a range of $\br$ with a fixed mass ratio $\bM=10$
and $\bbeta=0.2,0.5,1$. We also plot $U_0$ and $f_0$ with solid lines for comparison.
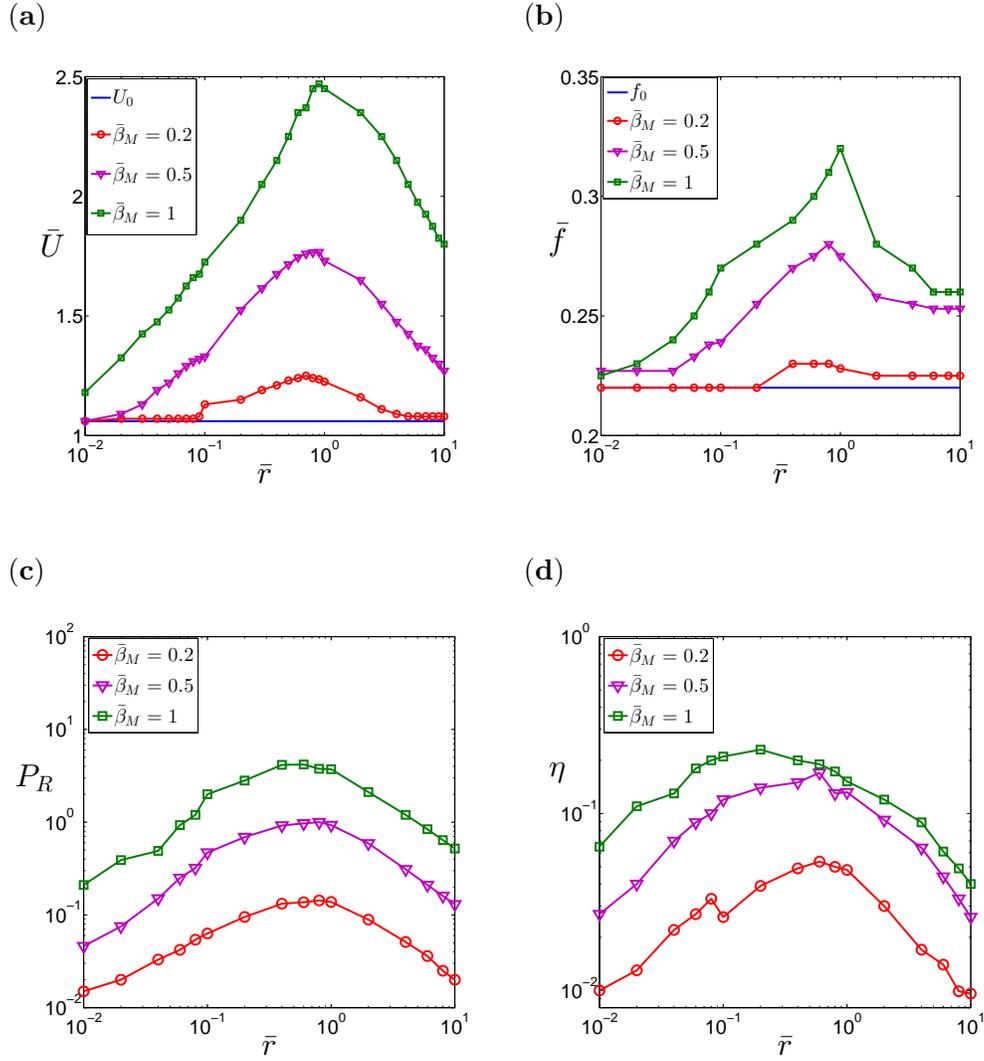
\Efourfigs{OptR}{OptRFreq}{OptRPower}{OptREfficiency}{\label{fig:optR}(a). Flutter velocity $\bU$ vs. resistance $\br$.
The solid line without symbols
denotes $U_0$, the flutter velocity when no piezoelectric material is included.
 (b). Vibration frequency right above the flutter boundary $\bar{f}$ vs. resistance $\br$.
The solid line without symbols
denotes $f_0$, the frequency when no piezoelectric material is included. (c). Averaged power $P_R$ vs. $\br$. (d). Averaged efficiency $\eta$ vs. $\br$. Other parameters used for three panels are $\bM=10$,
$\bbeta=0.2,0.5,1$.}{2.5}{2.5}

For all three $\bbeta$ values, the nondimensional resistance which leads to largest flutter speed is obtained around $\br=1$, 
when the fluid kinetic energy is balanced by the energy dissipated in the resistor and stored as electric energy in the piezoelectric material.
 The maximum flutter frequency and averaged electric power are also
achieved at these intermediate points. Since the efficiency scales with the inflow velocity as $\bU^{-3}$, the largest $\eta$ is obtained at a smaller value of $\br$ as
shown in panel (d). The exitance of an optimal resistance load was studied previously for piezoelectric beams in a vacuum with base excitations
\cite{dutoit2007experimental,erturk2009experimentally}, and was also observed for piezoelectric beams in viscous flow \cite{akcabay2012hydroelastic}.

Another important observation from figures \ref{fig:FlutterBdBeta} and \ref{fig:FlutterBdR} is that the numerical results show that the
increase in flutter speed by PZT (with increase in $\bbeta$ and $\br$) is more rapid for cases with higher $\bM$ ($\bM>1$), i.e.,
heavier flags in a lighter fluid. For lighter flags in a heavier fluid ($\bM\ll 1$), the piezoelectric material will have a negligible impact on the beam vibrational response for cases
with $\bbeta<0.2$ or $\br<0.1$.
\subsubsection{Piezoelectric effect with inductance}
Now we consider the effects of inductance on the flutter boundary. Similarly to the effect of the resistance, when inductance tends to zero ($\bl\rightarrow 0$), the circuit tends to a short circuit and the effect of the piezoelectric material is negligible. When inductance tends to infinity ($\bl\rightarrow\infty$), the circuit tends to an open circuit and the circuit is a purely resistive circuit as discussed in the previous section.

When the inductance has a moderate value, the parallel RL circuit has a natural frequency of 
$f_e=\displaystyle\frac{1}{2\pi\bl^{1/2}}$. 
In such cases, we find that the piezoelectric flag always exhibits a chaotic motion in the post-flutter regime.
 Periodic solutions with small perturbations are found close to the flutter boundary when $\bbeta<1$. 
Therefore, instead of considering the power and efficiency over one period, we use an averaged power $P_R$ and efficiency $\eta$ 
over a long time in the large-amplitude regime as measurements. In figure \ref{fig:FlutterL}, we show the critical flutter speed, vibration frequency and the
 corresponding averaged power and efficiency over a range of the circuit frequency $f_e$ with fixed
mass ratio $\bM=1$, $\br=1$ and various $\bbeta=0.2,0.5,0.8$.

\Efourfigs{M1inductor}{M1inductorFreq}{M1inductorEnergy}{M1inductorEfficiency}
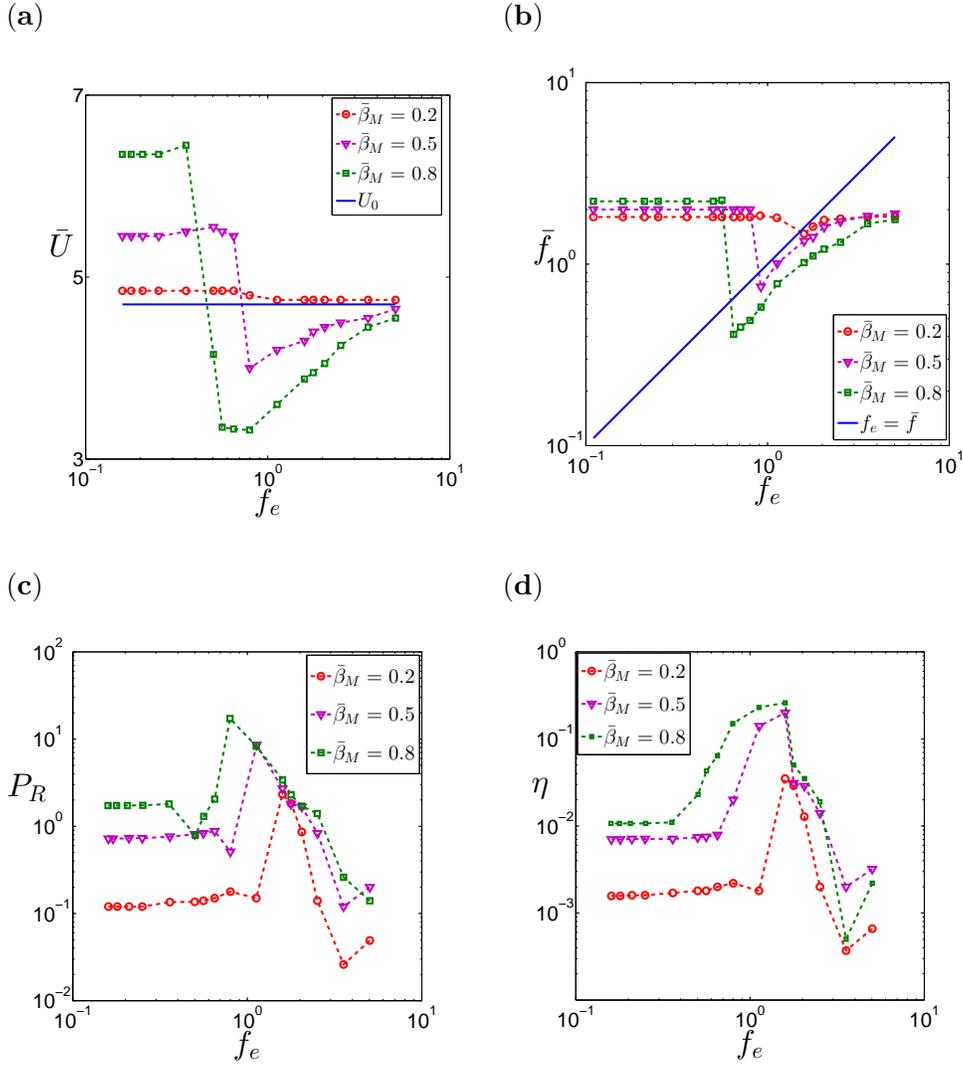
{\label{fig:FlutterL} Flutter boundary results for $\bM=1$, $\br=1$, and $\bbeta=0.2,0.5,0.8$. (a). Critical flutter speed $\bar{U}$ vs. circuit frequency
 $f_e=\displaystyle\frac{1}{2\pi\bl^{1/2}}$.
The solid line indicates the flutter speed $U_0$ when no piezoelectric material is included. (b). Vibration frequency $\bar{f}$ vs. circuit frequency $f_e$.
The solid line shows the line when $\bar{f}=f_e$, which corresponds to the ``lock-in'' situation. (c).
Averaged power dissipated on the resistance $P_R$ vs. $f_e$. (d). Averaged efficiency $\eta$ vs. $f_e$.}
{2.4}{2.4}
In panel (a) of figure \ref{fig:FlutterL}, we notice a drop in the critical flutter speed as the circuit frequency $f_e$ increases above a certain value, 
and the flutter speed after the drop is smaller as $\bbeta$ increases. The flag is able to flutter at a much lower speed due to the ``lock-in'' 
between the circuit resonant frequency and the flag resonant frequency, i.e. $\bfr=f_e$. In panel (b), we plot the vibration frequency $\bfr$ and the circuit frequency 
$f_e$ together with a solid line to denote the resonance situation where $\bfr=f_e$.
 We note that since the flag undergoes chaotic motion in this state, the oscillation frequency of the flag is not single-valued. We instead use 
the peak of the power spectrum of the frequency in the large-amplitude flapping state. 
Therefore, we observe some discrepancies between the flag vibration frequency and the circuit resonant frequency. Nonetheless, in a certain range of $\bl$ or $f_e$,
 the slope of the two frequency curves match well with each other, and the flutter speed decreases rapidly in the ``lock-in'' range.
This ``lock-in'' phenomenon was also observed by Xia \etal\cite{xia2015fluid}, and they also noticed a sudden decrease in the flutter speed with a slender body model.
 In the ``lock-in'' regime, we notice a rapid increase in the power dissipated in the
resistor, as well as the efficiency, as shown in panels (c) and (d). This is an important improvement for an energy harvester, since one of the limitations of energy harvesters is their requirement for flow speed which is difficult to achieve in a natural environment. By tuning the inductance into the ``lock-in'' regime, the energy harvester is able to flutter at a much lower speed and produce a much higher energy.

The flutter speed is greatly reduced in the ``lock-in'' regime as the flag jumps to a lower mode shape which requires less fluid kinetic energy to initiate flutter.
 If the flag is already in the lowest mode shape (second mode shape) and cannot be reduced further (such as for $\bM>1$ according to figure
\ref{fig:FlutterBdNoPiezo}(a)), we find that the flutter speed will not be 
reduced even though flag is in the ``lock-in'' condition. In figure \ref{fig:FlutterLMass}, we show the flutter speed and 
the corresponding vibration frequency over a range of circuit resonant frequencies with fixed values of $\bbeta$ and $\br$ and different mass ratios.
In figure \ref{fig:InductorModeShape}, we show the snapshots of the flag at different mass ratios with and without the inductor. The values of $\bl$ are chosen 
so that the vibration frequency of the beam is ``lock-in'' to the circuit resonant frequency. 

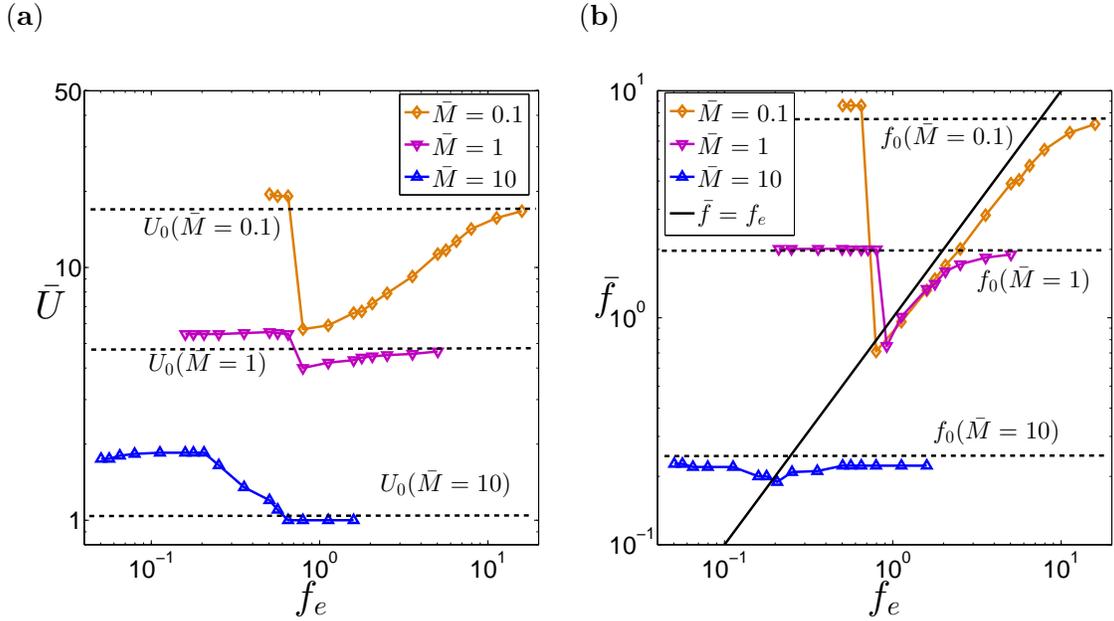
\Etwofigs{Minductor}{MinductorFreq}{\label{fig:FlutterLMass} (a) Flutter speed $\bU$ vs. $f_e=\displaystyle\frac{1}{2\pi\bl^{1/2}}$;
 (b) vibration frequency $\bfr$ vs. $f_e$ at $\bM=0.1,1,10$, $\bbeta=0.5$, and $\br=1$. 
Dashed lines denote the flutter speeds and frequency without piezoelectric effect for different mass ratios. Solid line in panel (b) denotes the curve $\bfr=f_e$.}{2.8}{2.8}
\Esixfigs{ModeM.0.1_U.5.8_beta.0.5_R.1_L0.04}{ModeM.1_U.4.2_beta.0.5_R.1_L0.02}{ModeM.10_U.1.9_beta.0.5_R.1_L0.6}
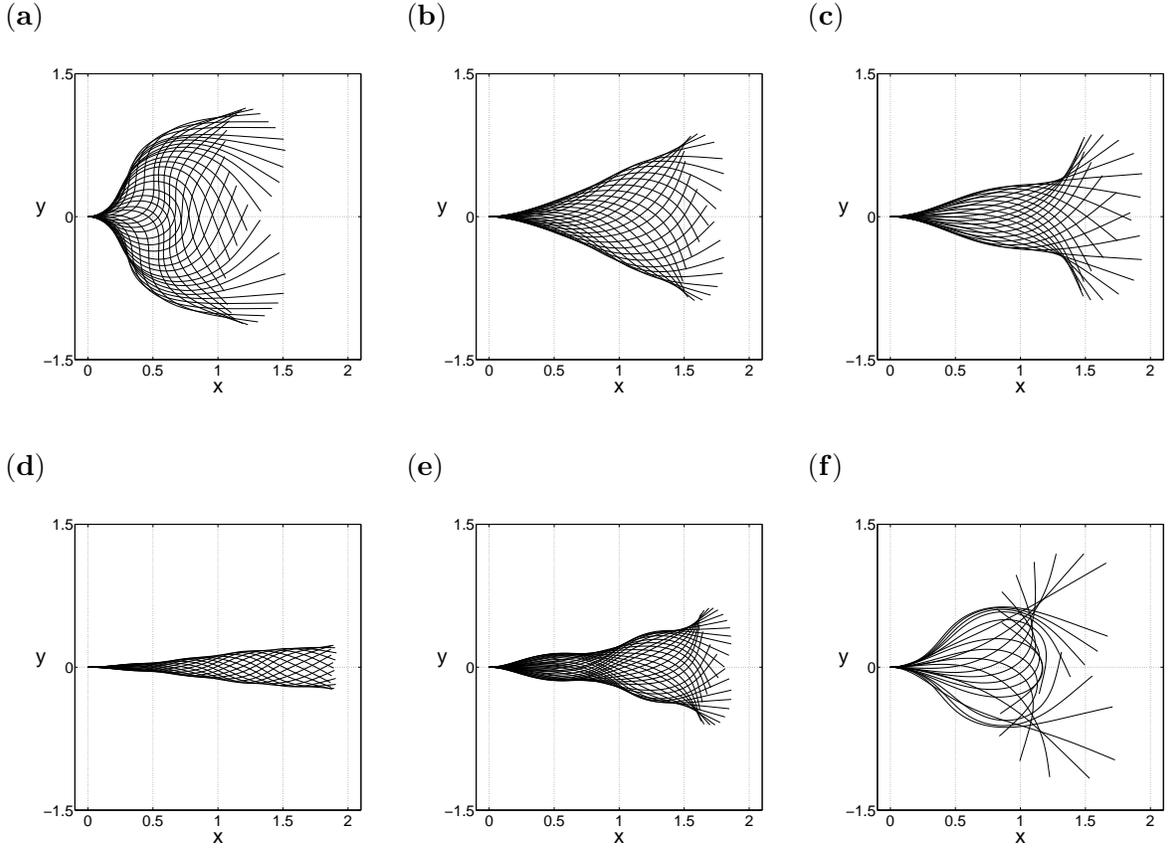
{ModeM.0.1_U.19.5_beta.0.5_R.1}{ModeM.1_U.5.6_beta.0.5_R.1}{ModeM.10_U.1.8_beta.0.5_R.1}{\label{fig:InductorModeShape}Flag snapshots at different $\bM$ 
and fixed values of $\bbeta=0.5$ and $\br=1$, with the inductor (a)-(c) at ``lock-in'' $\bfr=f_e$; and
without
the inductor (d)-(f). (a). Second mode shape, $\bM=0.1$, $\bU=5.8$, and $\bl=0.04$. (b). Second mode shape, $\bM=1$, $\bU=4.2$, and $\bl=0.02$.
(c). Second mode shape, $\bM=10$, $\bU=1.9$ and $\bl=0.6$. (d). Fifth mode shape, $\bM=0.1$ and $\bU=19.5$. (e) Third mode shape, $\bM=1$ and $\bU=5.6$.
(f). Second mode shape, $\bM=10$ and $\bU=1.9$. }{1.9}

Without the inductor, the flag flutters in the 5th and 3rd modes respectively for $\bM=0.1$ and 1, as shown in figures \ref{fig:InductorModeShape}(d) and (e). However,
in the ``lock-in'' regime, the flag flutters in the 2nd mode for both cases as shown in panel (a) and (b). When $f_e$ is small ($\bl\rightarrow 0$), the circuit
is close to a purely resistive circuit, and the flutter speed and vibration frequency with the PZT are larger than $U_0$ and $f_0$. Notice in figure 
\ref{fig:FlutterBdBeta}(a), the piezoelectric material is more effective for larger $\bM$ ($\bM>1$) at $\bbeta=0.5$. Therefore, the flutter velocity $\bU$
increases more for $\bM=10$ than the other two cases.  
As $\bl$ decreases or $f_e$ increases and approaches the lower bound of ``lock-in'' regime, the flutter speed and vibration frequency of the flag drop rapidly at first; as $f_e$
continues increasing, the flutter speed and vibration frquency gradually increases, eventually back to $U_0$ and $f_0$, as $\bar{l}\rightarrow 0$ or $f_e\rightarrow\infty$ 
corresponds to the short circuit scenario.  
When $\bM=10$, although the two frequencies still match each other around $f_e=0.2$ ($\bl\approx 0.6$), the flutter speed simply decreases to $U_0$ as $f_e$ increases.
In panel (c) and (f) for $\bM=10$, the flag flutters in the 2nd mode with/without the inductor, although the motions are different. In addition, notice in
figure \ref{fig:FlutterLMass} that the ``lock-in'' range is wider for lower $\bM$, where there is more room for the vibrating flag to drop down to the lowest
flutter mode (2nd mode). This indicates that when the fluid is light or the solid is heavy ($\bM>1$), 
the RL shunting circuit is not effective at modifying the flutter velocity or frequency. 
On the other hand, the results indicate that the RL shunting circuit can be used to tune the vibration responses of a lightweight piezoelectric flag 
in a heavy fluid such as water ($\bM\leq 1$).

\section{Scalability of piezoelectric beams}

Now, we consider several commonly used commercial materials and show how to apply the above discussion to actual devices in different fluids.
We first show some typical values of the above dimensionless parameters for PZT plates in air and water.
 We consider a rectangular plate with chord length of $0.05$ m, a thickness-to-chord ratio
of $0.5\%$, and a span-to-chord ratio
of 1. These are typical dimensions for a thin plate and are similar some of the experimental systems \cite{erturk2009experimentally,shelley2005heavy,watanabe2002experimental}.
PZT-5H is selected to be the piezoelectric material due to its high piezoelectric coupling coefficient relative to other PZT material.
Each PZT patch is assumed to have a fixed thickness $q_p$ of 0.02 mm, and the substrate material has a fixed thickness $q_s$ of 0.06 mm.
The dimensional piezoelectric coupling coefficient is then $6.6\times 10^{-5}$m$\cdot$ N/V, and
the internal capacitance is $6.4\times 10^{-6}$ F for each patch. The loaded resistance is chosen to be $1$ k$\Omega$ and the
inductance is $0.1$ H.

In Table \ref{tab:material}, we list the density, Young's modulus and Poisson's ratio for four common substrate materials
 (Balsa Wood, Polymer (POM), Aluminum Alloy and Steel) spanning a wide range of material density and elasticity values. 
We also list the material properties of the PZT-5H.
\begin{center}
\begin{table}[ht]
\caption{\label{tab:material}Parameters of the Solid Material and PZT-5H}
\begin{tabular}{|c|>{\centering\arraybackslash}p{2cm}|>{\centering\arraybackslash}p{2cm}|>{\centering\arraybackslash}p{2cm}|>{\centering\arraybackslash}p{2cm}|>{\centering\arraybackslash}p{2cm}|}
\hline\hline
  &Balsa Wood & POM & Alumina Alloy &Steel &PZT-5H\\
\hline
Density $\rho$ $(\mbox{kg}/\mbox{m}^3)$&97.1&1480&2800&7850&7700\\
\hline
Young's Modulus $E$ (GPa)&1.8&3&79&210&66\\
\hline
Poisson's Ratio&0.36&0.35&0.33&0.3&0.34\\
\hline\hline
\end{tabular}
\end{table}
\end{center}
The bending rigidity for the composite piezoelectric flag is calculated by:
\bq
 B=EI=\displaystyle\frac{E_s}{1-\nu_s^2}\displaystyle\frac{q_s^3S}{12}+\frac{E_p}{1-\nu_p^2}S(\frac{1}{2}q_s^2q_p+q_sq_p^2+\frac{2}{3}q_p^3)
\eq
 where $E_s, \nu_s$ and $E_p,\nu_p$
are the Young's modulus and Poisson's ratio for the substrate and the PZT patch, respectively. In figure \ref{fig:nondimparameter}, we show the dimensionless $\bar{M}$, $\bbeta$
and $\br$ for different materials according to the definitions given in Table \ref{tab:nondimen}.
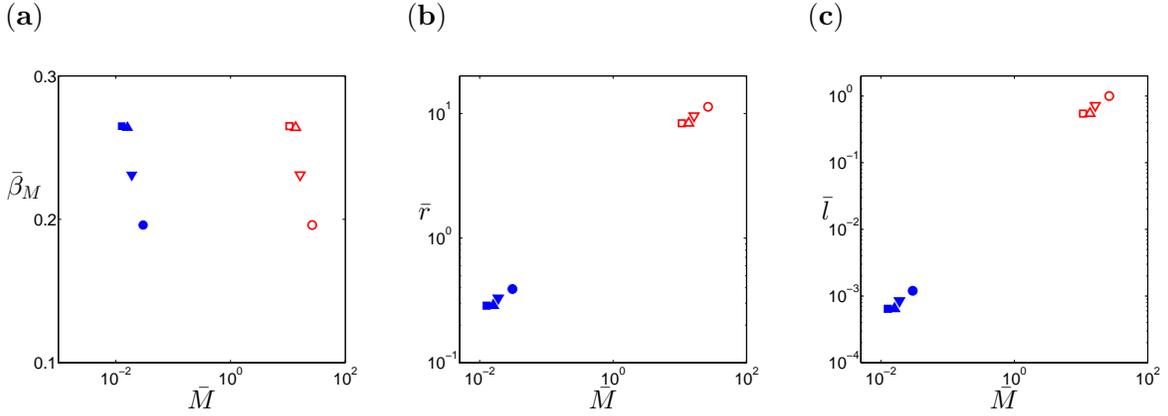
\Ethreefigs{nondimparameter}{nondimparameter2}{nondimparameter3}{\label{fig:nondimparameter} Range of dimensionless parameters corresponding to different materials.
The open symbols indicate materials in air, and the filled symbols indicate materials in water. $\circ/\triangledown/\triangle/\square$:
steel/Aluminum alloy/POM/Balsa wood in air; $\bullet/\blacktriangledown/\blacktriangle/\blacksquare$: steel/Aluminum alloy/POM/Balsa wood
in water.
(a). $\bM$ vs. $\bbeta$; (b). $\bM$ vs. $\br$; (c). $\bM$ vs $\bl$. }{1.9}{1.9}{1.9}
We consider two types of fluids: air and water to cover the range of $\bM$, i.e., heavy flags in light fluid and light flags in heavy fluid, respectively.

First, we note that
$\bbeta$ is unchanged as the fluid medium changes from air to water since it is not a function of $\rho_f$, as shown in Table \ref{tab:nondimen}.
 Notice from Table \ref{tab:material}, the PZT patch has a large density and Young's modulus compared to many substrate materials.
In general, the mass ratio of various flags in water is much smaller than those in air,
$\bM$ is around $10-30$ for flags in air, and therefore they mainly flutter in the second mode. For flags in water, $\bM$ is around $0.01-0.03$; they flutter in a 
higher mode and the critical flutter speed without PZT is much larger, as shown in figure \ref{fig:FlutterBdNoPiezo}.

The dimensionless resistance density $\br$ and inductance density $\bl$ both decrease substantially
when the fluid becomes heavier as shown in panels (b) and (c) of figure \ref{fig:nondimparameter}. We also note that
the dimensionless $\br$ and $\bl$ depend linearly on the values of resistance and inductance, and therefore can be easily adjusted by changing their values.

Now, we take POM as an example and show the dimensional flutter speed, vibration frequency, averaged power and efficiency in air and water under 
different conditions in Table \ref{tab:scale}. We consider both the resistance only circuit and the RL circuit. 
The effectiveness of the resistance only circuit on modifying the flutter speed and frequency is larger for flags in air,
 and practically negligible for flags in water. By using the optimal resistance value in a resistance only circuit shown in Table \ref{tab:scale}, 
the flutter speed is increased by $50\%$ compared to the case without PZT when the flag is placed in air, 
but only by $2\%$ when it is placed in water. A larger electric power and efficiency are also obtained with the optimal R when the flag is placed in air.
For heavy fluids, the optimal R value which leads to a larger flutter velocity results in a smaller
power and efficiency. The voltages for different R values are almost identical in this case, so the power and efficiency decrease with a higher resistance.  

On the other hand, for an RL shunting circuit, the flag in the lighter fluid mainly flutters in the second mode (the lowest flutter mode), 
and hence the flutter speed and the frequency will not be greatly reduced by the inductance, and there is not an optimal inductance value. 
The RL circuit is more effective for lighter flags in a heavier fluid, 
as shown in the example of water in Table \ref{tab:scale}. We consider the RL circuit with the optimal R value. 
When the flag is placed in water, by tuning the inductor to the ``lock-in'' range, the flutter 
speed is reduced by $75\%$ compared to the case without PZT. The averaged power increased by a factor of 10 compared to the case with a purely resistive circuit and 
the efficiency increased by a factor of 1000. Note that both circuits have important practical applications. 
An increase of the flutter speed is typically desired for vibration control applications, while an decrease 
is typically desired for energy harvesting applications. 

\begin{table}[ht]
\caption{\label{tab:scale}The effect of the PZT on the flutter speed, frequency, power and efficiency for a POM flag in air and water}
\begin{tabular}{|c|>{\centering\arraybackslash}p{2cm}>{\centering\arraybackslash}p{2cm}>{\centering\arraybackslash}p{2cm}|>{\centering\arraybackslash}
p{2cm}>{\centering\arraybackslash}p{2cm}>{\centering\arraybackslash}p{2cm}|}
\hline\hline
Fluid&\multicolumn{3}{|c|}{Air ($\rho_f=1.18$ kg/m$^3$)} &\multicolumn{3}{|c|}{Water ($\rho_f=998$ kg/m$^3$)}\\
&\multicolumn{3}{|c|}{$\bM=13.6,\bbeta=0.26$}&\multicolumn{3}{|c|}{$\bM=0.016,\bbeta=0.26$}\\
\hline
Resistance only & No PZT&R&Optimal R& No PZT&R&Optimal R\\
circuit& & 1000$\Omega$&127$\Omega$&&1000$\Omega$&3477$\Omega$\\
\hline
Flutter speed (m/s)&14.9&20&22.4&18&18.2&18.4\\
Frequency (Hz)&131&137&118&323&150&150\\
Power $P_R$ (W)&-&0.67&3.55&-&0.93&0.27\\
Efficiency $\eta$&-&0.023&0.078&-&3.4e-4&1.0e-4\\
\hline
RL circuit & No PZT& L & No Optimal &No PZT&L&Optimal L\\
&-&0.1H& L &&0.1H&0.93H\\
\hline
Flutter speed (m/s)&14.9&14.9&-&18&8.7&4.9\\
Frequency (Hz)&131&131&-&323&44&13\\
Power $P_R$ (W)&-&0.32&-&-&1.78&3.1\\
Efficiency $\eta$&-&0.0015&-&-&0.02&0.14\\
\hline\hline
\end{tabular}
\end{table}

For an actual device with a larger size and complete geometric similarities, i.e., 
if the thickness-to-chord ratio, thickness ratio of the PZT patch to substrate material, and span-to-chord ratio are fixed, 
then the dimensionless $\bM$, $\bbeta$ and $\br$ are unchanged with the length scale for the same PZT and substrate material, fluid and resistance values. 
According to Table \ref{tab:nondimen}, $\bbeta=\displaystyle\beta_M\sqrt{\frac{1}{Bc_p}}$. If we assume the length scale is $b$, then 
the dimensional $\beta_M$ scales as $b^2$, $B$ scales as $b^4$, and capacitance density is unchanged with the length, which implies that $\bbeta$ is unchanged
with the length scale. A similar analysis shows that $\br$ is also invariant with the length. 
$\bl$ is proportional to $b$, but when the length increases, the same dimensionless value can be obtained by changing the value of the inductor.
Therefore, the dimensionless results obtained from the model problems can be directly applied to the full scale problems. The dimensional velocity
is invariant with respect to $b$, the frequency scales as $b^{-1}$ and the power scales as $b^2$. Therefore, when the problem is extended to a larger size, the flutter 
speed is unchanged, but a smaller frequency and a larger power are expected.

On the other hand, the actual thickness of the PZT patch can not exceed several centimeters due to material properties and manufacture restrictions. 
Thus, the thickness ratio of the PZT patch and the substrate material can be much smaller than that of the model scale problem, which leads to a smaller $\bbeta$. 
Another important issue is that the thickness to chord ratio is generally larger for an actual device, otherwise the plate may undergo material failure, which
is not considered in the current analysis. The bending rigidity $B$ is larger for a thicker plate, and it leads to a larger flutter speed which scales as $B^{1/2}$ according to Table \ref{tab:nondimen}.
Moreover, although the substrate material can be scaled, the material properties of the PZT patch can be difficult to scale. For example, very large PZT patches
 are generally not feasible in reality, so several discrete patches are often linked together as an alternative. 
These issues along with the stability of the particular material in a certain fluid must be considered when extending the model scale results to full scale problems.
\section{Conclusions}
In this work, we investigate the effect of the PZT material on the flutter speed, vibration mode and frequency, and energy harvesting power and efficiency of a flexible flag
 in different fluid medium. We develop a fully coupled fluid-solid-electric model by combining the inviscid vortex sheet model with a
linear electro-mechanical coupling model. A cantilever flag with two layers of PZT and a uniform inflow is studied as a model problem. 
Based on a nondimensional analysis, four dimensionless parameters $\bM,\bbeta,\br,\bl$ were found to govern the reduced flutter velocity $\bU$ and frequency $\bfr$ of the piezoelectric flag.

When the mass ratio $\bM$ is small, the flag flutters at higher mode, as the relative high fluid inertia and fluid disturbing force allow higher
wave number deforming patterns. Flags with lower $\bM$ also require a larger inflow velocity to initiate flutter, due to the higher relative fluid
inertia and the damping resistance against motion. 

Two types of output circuits are considered in this work: a purely resistive circuit and a resonant RL circuit. For a purely resistive shunting circuit,
 the PZT material functions as a stabilizer to the system by increasing the effective system stiffness and damping. Therefore, as $\bbeta$ increases, 
the flutter speed, as well as the vibration frequency, averaged power and efficiency also increase. In particular, we find that the power scales as $\bbeta^2$.
For a resistance only circuit, the optimal resistance to maximize the flutter speed and the power is obtained at near $\br=1$, when the fluid kinetic energy is balanced by the energy dissipated in 
the resistor and stored as electrical energy in the piezoelectric material. The results also show that a purely resistive circuit have a greater impact for heavier flags
in lighter fluids, i.e., $\bM>1$. For $\bM\ll 1$, the PZT flag with a resistance only circuit has a negligible impact if $\bbeta<0.2$ and/or $\br<0.1$, which is the case
for most PZT flags in water.  
For a resonant RL circuit, an optimal inductor is found when the circuit frequency matches the flag's vibration frequency.
 In this ``lock-in'' regime, the flutter speed and frequency are greatly reduced and the flag will flutter in a lower mode. 
The ``lock-in'' range is found to be larger, and the flutter speed and the vibration frequency drop more in this range for lower $\bM$. 
Maxima electric power and efficiency are also obtained in the ``lock-in'' regime. 

We also consider a model scale set-up with several commonly used commercial material in air and water. 
Typical ranges of dimensionless parameters are obtained for four types of materials that spans a wide range of solid density and rigidity values. 
For POM, we find that the resistance only circuit is more effective when the flag is placed in a lighter fluid like air. An RL circuit is able to 
significantly reduce the flutter speed and frequency when the flag is placed in a heavier fluid like water, but is not as effective for similar flags in air.
 For vibration control
applications, an increase in the flutter speed is typically desired. On the other hand, for energy harvesting applications, a lower flutter speed is desired.

An extension of this work is to consider the application of the PZT material on passive vibration control and for energy harvesting with a vibrating foil, 
which requires generalizing the current vortex sheet model to more complicated geometric shapes, and considering a combined bending and twisting deformation. 
Moreover, the influence of viscous and 3-D effects on electro-fluid-solid interactions requires additional study. 
\begin{acknowledgments}
We acknowledge support from a Sloan Research Fellowship and NSF-DMS 1329726 (S.A.). We are also grateful to the funding provided by Office 
of Naval Research (ONR) grant no. N00014-13-0383, managed by Dr. Ki-Han Kim.
\end{acknowledgments}
\bibliographystyle{unsrt}

\end{document}